\documentclass{article}
\usepackage[LGR,T1]{fontenc}
\usepackage{amsmath}
\usepackage{amssymb}
\usepackage{graphicx}
\usepackage{subcaption}

\makeatletter

\ProvideTextCommand{\~}{LGR}[1]{\char126#1}

\DeclareTextSymbolDefault{\textquotedbl}{T1}

\makeatother

\begin{document}
\thispagestyle{empty}
\vspace*{-15mm}

\begin{flushleft}
{\bf OUJ-FTC-4}\\
{\bf OCHA-PP-358}\\
\end{flushleft}

\vspace{8pt}
\begin{center}
{\Large\bf
Beyond Squeezing \`a la Virasoro Algebra
}
\vspace{7pt}
\baselineskip 36pt

\renewcommand{\thefootnote}{\fnsymbol{footnote}}

So Katagiri${}^{1}$\footnote[0]{So.Katagiri@gmail.com}, Akio Sugamoto${}^{2, 3}$, Koichiro Yamaguchi${}^{4}$, Tsukasa Yumibayashi${}^{5}$

\vspace{32pt}

\small{\it ${}^{1}$Division of Arts and Sciences, The School of Graduate Studies,
The Open University of Japan, Chiba 261-8586, Japan\\
${}^{2}$Tokyo Bunkyo Study Center, The Open University of Japan (OUJ), \\
Tokyo 112-0012, Japan \\
${}^{3}$Ochanomizu University, 2-1-1 Ohtsuka, Bunkyo-ku, Tokyo 112-8610, Japan\\
${}^{4}$Nature and Environment, Faculty of Liberal Arts, The Open University of Japan, Chiba 261-8586, Japan\\
${}^{5}$Department of Social Information Studies, Otsuma Women's University, 12 Sanban-cho, Chiyoda-ku, Tokyo 102-8357, Japan}

\end{center}

\vspace{10pt}

\begin{center}

\baselineskip 16pt
\noindent

\begin{abstract}
The generalization of squeezing is realized in terms
of the Virasoro algebra. The higher-order squeezing
can be introduced through the higher-order time-dependent potential,
in which the standard squeezing operator is generalized
to higher-order Virasoro operators. We give a formula that
describes the number of particles generated by the higher-order squeezing
when a parameter specifying the degree of squeezing is small.
The formula (\ref{eq:NsqNum}) shows that the higher the order of squeezing becomes the larger
the number of generated particles grows.
\end{abstract}

\end{center}

\newpage

\section{Introduction}

Squeezing has received great attention
 in various fields such as quantum
optics \cite{key-9}, cosmology \cite{key-10} 
and quantum information \cite{key-12}.

In particular, particle generation is an important issue; in the time-dependent
oscillator, the number of produced particles is increased by repeated
application of squeezing.

Accordingly, it is expected that a generalization of squeezing can increase 
the number of the generated particles more effectively.

Braunstein and McLachlan \cite{key-7} generalized the parametric amplification
by producing k-photon correlation, and numerically showed the structure
of phase space. There, the usual quadratic squeezing was extended
to the cases with cubic, quartic or higher-order interactions. Next,
the statistics of heterodyne and homodyne detection in cubic and quartic
interactions were examined by Braunstein and Caves \cite{key-8}. These
studies have shown that the features of higher-order interactions
are well reflected in the structure of the phase space. As a review, see \cite{key-6}.

In this paper, we investigate the generalized squeezing from the viewpoint
of the Virasoro algebra.

The Virasoro algebra is an algebra used in string theory and represents
symmetry including scale transformation \cite{key-3} which is called the
conformal symmetry. The symmetries of parallel translation and scale transformation
make up $\mathrm{SL}(2,\mathbb{C})$ as a subalgebra of the Virasoro algebra.

Squeezing is a transformation of phase space, which preserves
the area of the phase space of position and momentum. By the squeezing
transformation, the position is scaled up (down) and the momentum
is scaled down (up) in the opposite direction. Therefore, squeezing
can be considered as a kind of scale transformation combined
with parallel translation to form subalgebra of Virasoro algebra.
This is the standpoint from which we construct a theory in this study.

The paper is organized as follows. First, in the next section,
we introduce the Virasoro algebra with conformal symmetries. Next, in section 3,
we examine the usual second-order squeezing
in light of time-dependent oscillators.

In section 4, we investigate the N-th order squeezing and show that this
satisfies the Virasoro algebra. In section 5, we discuss the particle
generation of the N-th order squeezing. In section 6, we discuss the uncertainty relation of the N-th order squeezing.
We examine the structure of the phase space of N-th order squeezing in section
7. The final section is devoted to discussion.

\section{Virasoro algebra}

In this section, we introduce the Virasoro algebra and see that it
contains a scale transformation.
Virasoro algebra is an algebra generated by generators $L_n$
that satisfy the following algebraic relation,

\begin{equation*}
[L_{n},L_{m}]=(n-m)L_{n+m}+\frac{c}{12}(n^{3}-n)\delta_{m+n,0},
\end{equation*}
where, $c$ is called the central charge and commutes with any $L_{n},\ [c,L_{n}]=0.$

$L_{-1},L_0$ and $L_{-1}$ satisfies 
\begin{equation*}
    [L_{\pm1},L_0]=\pm L_{\pm1},\ [L_1, L_{-1}] = 2L_{0} 
\end{equation*}
which is $\mathrm{SL}(2,\mathbb{C})$ subalgebra of Virasoro Algebra.

When $c=0$, this algebra reduces to

\begin{equation*}
[L_{n},L_{m}]=(n-m)L_{n+m},
\end{equation*}
which is called Witt algebra
(centerless Virasoro algebra) \cite{key-4}.

The generator of this algebra $L_{n}$ can be constructed by $z$
and its differential operator $\partial$ as follows:

\begin{equation*}
L_{n}=z^{n+1}\partial.
\end{equation*}

Then, we can understand the specific geometrical meaning of these
algebras.

\begin{equation*}
L_{0}=z\partial
\end{equation*}
generates a scale transformation,
\begin{equation*}
e^{\theta L_{0}}f(z)=f(e^{\theta}z).
\end{equation*}

Next, 
\begin{equation*}
L_{-1}=\partial
\end{equation*}
generates a parallel transformation,
\begin{equation*}
e^{\theta L_{-1}}f(z)=f(z+\theta).
\end{equation*}

Finallly, 
\begin{equation*}
    L_{1}=z^2\partial
\end{equation*}
generates a special conformal transformation,
\begin{equation*}
    e^{\theta L_{1}}f(z)=f\left(\frac{z}{1-\theta z}\right).
\end{equation*}

Now, the Virasoro algebra (or Witt algebra)
contains scale transformation and parallel translation in it.

\section{Squeezing and time-dependent oscillators}

In this section, we introduce the usual second-order squeezing, and in particular,
show that the squeezed state can be obtained from time-dependent oscillators.

As is well known, in a system that describes harmonic oscillators,
squeezed states $|\theta\rangle$ are constructed by acting on the
states with the following operator,

\begin{equation*}
e^{\theta \hat{F}},\ \hat{F}=\frac{\hat{a}^{2}-\hat{a}^{\dagger2}}{2},
\end{equation*}
where $\theta$ is a squeezing parameter, and $\hat{a}^{\dagger}$ and
$\hat{a}$ are a creation and annihilation operator of the harmonic oscillator.
The position and momentum operator are constructed as follows from $\hat{a}^{\dagger}$
and $\hat{a}$,
\begin{align*}
    \left\{\ 
    \begin{aligned}
    \hat{x}=\sqrt{\frac{1}{2\omega_{0}}}(\hat{a}+\hat{a}^{\dagger}), \\
    \hat{p}=i\sqrt{\frac{\omega_{0}}{2}}(\hat{a}-\hat{a}^{\dagger}).
    \end{aligned}
    \right.
\end{align*}

We note that the generator $\hat{F}$ is rewritten into the following symmetric
form using $\hat{x}$ and $\hat{p}$,
\begin{equation*}
\hat{F}=\frac{i}{2}\left(\hat{x}\hat{p}+\hat{p}\hat{x}\right).
\end{equation*}

Therefore, $\hat{x}$ and $\hat{p}$ are scaled by $\hat{F}$ as follows,

\begin{align*}
    \left\{\ 
    \begin{aligned}
        e^{\theta \hat{F}}\hat{x}e^{-\theta \hat{F}}&=e^{\theta}\hat{x}, \\
        e^{\theta \hat{F}}\hat{p}e^{-\theta \hat{F}}&=e^{-\theta}\hat{p}.
    \end{aligned}
    \right.
\end{align*}

The scaling factor of $\hat{x}$ and $\hat{p}$ are inverse with one another because $\hat{F}$ is symmetric while the commutation relation antisymmetic i.e. $[\hat{x},\hat{p}]=-[\hat{p},\hat{x}]$.

These relation reduce that the squeezed state
preserves the minimum uncertainty relation,
\begin{equation*}
\langle\theta|\Delta \hat{x}^{2}|\theta\rangle\langle\theta|\Delta \hat{p}^{2}|\theta\rangle=\frac{1}{4}\left|\langle\theta|[\hat{x},\hat{p}]|\theta\rangle\right|^{2}=\frac{1}{4},
\end{equation*}
where
\begin{align*}
    &\left\{ \ 
    \begin{aligned}
        \Delta \hat{x} &=\hat{x}-\langle\theta|\hat{x}|\theta\rangle=&&\hat{x}-e^{\theta}\langle0|\hat{x}|0\rangle, \\
        \Delta \hat{p} &=\hat{p}-\langle\theta|\hat{p}|\theta\rangle=&&\hat{p}-e^{-\theta}\langle0|\hat{p}|0\rangle,
    \end{aligned}
    \right. \\
    &\left\{ \ 
    \begin{aligned}
        \langle\theta|\Delta \hat{x}^{2}|\theta\rangle&=e^{2\theta}\left(\langle0|\hat{x}^{2}|0\rangle-\langle0|\hat{x}|0\rangle^{2}\right)=&&\frac{1}{2}e^{2\theta}, \\
        \langle\theta|\Delta \hat{p}^{2}|\theta\rangle&=e^{-2\theta}\left(\langle0|\hat{p}^{2}|0\rangle-\langle0|\hat{p}|0\rangle^{2}\right)=&&\frac{1}{2}e^{-2\theta}.
    \end{aligned}
    \right.
\end{align*}

The squeezing opertor can be rotated to any direction, namely,

\begin{equation*}
\hat{F}(\phi)=\frac{i}{2}\left(\hat{x}(\phi)\hat{p}(\phi)+\hat{p}(\phi)\hat{x}(\phi)\right),
\end{equation*}
where
\begin{align*}
    \left\{ \ 
    \begin{aligned}
        \hat{x}(\phi)&= \hat{x}\cos\phi-\hat{p}\sin\phi, \\
        \hat{p}(\phi)&=\hat{p}\cos\phi+\hat{x}\sin\phi.
    \end{aligned}
    \right.
\end{align*}

The time-dependent oscillators are described by,

\begin{equation*}
\hat{H}=\hat{H}_{0}+\frac{1}{2}\omega(t)\hat{x}^{2},
\end{equation*}
where
\begin{equation*}
\hat{H}_{0}=\frac{\hat{p}^{2}}{2}+\frac{1}{2}\omega_{0}\hat{x}^{2}.
\end{equation*}
In the interaction picture, the Hamiltonian can be written as,

\begin{equation*}
\hat{H}_{I}=\frac{1}{2}\omega(t)\hat{x}_{I}^{2},
\end{equation*}
where $\hat{x}_{I}=e^{i\hat{H}_{0}t}\hat{x}e^{-i\hat{H}_{0}t}=\hat{x}(\omega_{0}t)$. The time evolution
of the state can be described as follows,
\begin{equation*}
|0(t)\rangle=\hat{U}(t,0)|0\rangle,\ \hat{U}(t,0)=Te^{i\int_{0}^{t}dt\frac{1}{2}\omega(t)\hat{x}_{I}^{2}},
\end{equation*}
where $T$ is the time ordering operator.

If we divide the time evolution operator into the product of those
in a small time lapse and write,

\begin{equation*}
\hat{U}(t,0)|0\rangle=\hat{U}(t,n\Delta t)\cdots \hat{U}(\Delta t,0)|0\rangle,
\end{equation*}
we can write the infinitesimal time evolution operator $\hat{U}(t+\Delta t,t)$
becomes,
\begin{equation*}
\hat{U}((n+1)\Delta t,n\Delta t)=Te^{[-i\frac{1}{2}\alpha(t)\hat{x}_{I}^{2}(t)]_{n\Delta t}^{(n+1)\Delta t}-i\int_{n\Delta t}^{(n+1)\Delta t}\alpha(t)\left\{ \hat{x}_{I}(t)\hat{p}_{I}(t)+\hat{p}_{I}(t)\hat{x}_{I}(t)\right\} },
\end{equation*}
where $\alpha$ is given by $\dot{\alpha}(t)=\omega(t)$.
If $\alpha(t)=\sum_{n}\delta(t-(n+\frac{1}{2})\Delta t)$ , we obtain
\begin{equation*}
\hat{U}(t,0)|0\rangle=e^{\theta_{n}\hat{F}_{n}}\cdots e^{\theta_{0}\hat{F}_{0}}|0\rangle,
\end{equation*}
where $\theta_{n}=2(n+\frac{1}{2})\Delta t,\ \hat{F}_{n}=\hat{F}(2(n+\frac{1}{2})\Delta t)$.
Then, in time-dependent oscillators, the time evolution of the state
can be described as the repetition of the squeezing operations. 

The above description is crucial in the generalization of the squeezing phenomenon in the following sections.

\section{N-th order squeezing}
In this section, N-th order squeezed state is represented by the Virasoro algebra, i.e. Witt algebra,
 as the time dependent anharmonic oscillator.
\subsection{N-th order squeezing}
We can construct the Virasoro algebra from the position and momentum operators
in quantum mechanics\footnote{We take $\hbar$ = 1. Here we have tentatively asssumed the operator
ordering in defining (\ref{eq:Ln}). The more detailed analysis
may give a non-vanishing central charge, but whether the analysis
restores the violation of the minimum uncertainty relation for the
N-th order squeezed state or not, is not clear at this moment. (See the
Discussion.)},

\begin{equation}
\hat{L}_{n}\equiv-\frac{i}{2}\left(\hat{x}^{n+1}\hat{p}+\hat{p}\hat{x}^{n+1}\right).\label{eq:Ln}
\end{equation}
$\hat{L}_{n}$ satisfies
\begin{equation*}
[\hat{L}_{n},\hat{L}_{m}]=(n-m)\hat{L}_{n+m},
\end{equation*}
This is the centerless Virasoro algebra (Witt algebra).\footnote{Here the calculation has been performed using a
position representation, $p=-i\frac{\partial}{\partial x}$. This
calculation can be extended to $n\in\mathbb{C}$.} Thus, $\hat{L}_{n}$ represents a conformal transformation
in $x$ space. We note here that $\hat{L}_{0}$ is a generator of the usual second-order squeezing transformation.

Similarly, the dual operator of $\hat{L}_{n}$,

\begin{equation}
\hat{\tilde{L}}_{n}\equiv\frac{i}{2}\left(\hat{p}^{n+1}\hat{x}+\hat{x}\hat{p}^{n+1}\right)
\end{equation}
also satisfies the Virasoro algebra,
\begin{equation*}
[\hat{\tilde{L}}_{n},\hat{\tilde{L}}_{m}]=(n-m)\hat{\tilde{L}}_{n+m}.
\end{equation*}

This shows that $\hat{\tilde{L}}_{n}$ represents a conformal transformation
in $p$ space.

If we express $\hat{L}_{n}$ as a harmonic oscillator,
it has the following form,
\begin{equation*}
\hat{L}_{n}=\frac{i}{2}\left(-i\sqrt{\frac{\omega_{0}}{2}}\right)\left(\sqrt{\frac{1}{2\omega_{0}}}\right)^{n}\left((\hat{a}+\hat{a}^{\dagger})^{n}(\hat{a}-\hat{a}^{\dagger})+(\hat{a}-\hat{a}^{\dagger})(\hat{a}+\hat{a}^{\dagger})^{n}\right).
\end{equation*}

We call the state, obtained by applying the unitary
operator $\hat{L}_{n}$ to the vacuum, an ``(Virasoro)
N-th order squeezed state'' and write it as follows,
\begin{equation}
|\theta\rangle_{n}\equiv e^{\theta L_{n}}|0\rangle.
\end{equation}

Let us examine how $\hat{x}$ and $\hat{p}$ are transformed by the N-th order squeezing.
First, 
\begin{equation*}
[\hat{L}_{n},\hat{x}]=-\hat{x}^{n+1}
\end{equation*}
leads to the transformation of $x$
\begin{align}
& S_{n}\hat{x}S_{n}^{\dagger} =\sum_{k=0}^{\infty}A_{k}(\theta \hat{x}^{n})^{k}\hat{x},\\
& \left\{
\begin{aligned}
 & A_{0}=1,\\
 & A_{1}=-1,\\
 & {A_{k}=(-1)^{k}\frac{\prod_{j=1}^{k-1}(jn+1)}{k!}},\ k>1,
\end{aligned}
\right. \notag
\end{align}
where $\hat{S}_{n}=e^{\theta \hat{L}_{n}}$.
The sum can be estimated as
\begin{equation}
S_{n}xS_{n}^{\dagger}=(1+n(\theta x^{n}))^{-\frac{1}{n}}x.
\end{equation}
If $n=0$, this equation is reduced to the usual second-order squeezing,
\begin{equation*}
\hat{S}_{0}\hat{x}\hat{S}_{0}^{\dagger}=e^{-\theta}\hat{x}.
\end{equation*}
Next, 
\begin{equation*}
[\hat{L}_{n},\hat{p}]=i(1+n)\hat{L}_{n-1},
\end{equation*}
leads to the transformation of $\hat{p}$
\begin{align}
&\hat{S}_{n}\hat{p}\hat{S}_{n}^{\dagger}  =\sum_{k=0}^{\infty}B_{k}\theta^{k}\hat{L}_{kn-1},\\
& \left\{
\begin{aligned}
 & B_{0}=1,\\
 & B_{1}=(1+n),\\
 & B_{k}=\frac{\prod_{j=1}^{k}(1-(j-2)n)}{k!}.
\end{aligned}
\right. \notag
\end{align}
This sum can also be calculated as
\begin{equation}
\hat{S}_{n}\hat{p}\hat{S}_{n}^{\dagger}=\frac{(1+n(\theta \hat{x}^{n}))^{\frac{1}{n}+1}\hat{p}+\hat{p}(1+n(\theta \hat{x}^{n}))^{\frac{1}{n}+1}}{2}.\label{eq:SPS}
\end{equation}
If we take the commutative limit, that is, $\hat{x}$ and $\hat{p}$ commute
in (\ref{eq:SPS}), these equations yield
\begin{align}
    \left\{ \ 
    \begin{aligned}
        &\hat{S}_{n}\hat{x}S_{n}^{\dagger}=(1+n(\theta \hat{x}^{n}))^{-\frac{1}{n}}\hat{x}, \\
        &\hat{S}_{n}\hat{p}\hat{S}_{n}^{\dagger}=(1+n(\theta \hat{x}^{n}))^{\frac{1}{n}+1}\hat{p}.
    \end{aligned}
    \right.
\end{align}

In this limit, the phase space volume is multiplied
by $(1+n(\theta \hat{x}^{n}))$ by N-th order squeezing.

Using the $\hat{x}$ and $\hat{p}$ operators rotated by angle $\phi$,
\begin{align*}
    \left\{ \ 
    \begin{aligned}
        \hat{x}(\phi)=\hat{x}\cos\phi-\hat{p}\sin\phi,\\
        \hat{p}(\phi)=\hat{x}\sin\phi+\hat{p}\cos\phi,
    \end{aligned}
    \right.
\end{align*}
the generalization of $\hat{L}_{n}$ in any direction, as in the usual second-order squeezing,
is naturally defined by
\begin{equation}
\hat{L}(\phi)_{n}=-\frac{i}{2}\left(\hat{x}^{n+1}(\phi)\hat{p}(\phi)+\hat{p}(\phi)\hat{x}^{n+1}(\phi)\right).
\end{equation}
Then, $\hat{L}(\phi)$ can be expanded in terms of $x$and p as follows:
\begin{align}
\hat{L}(\phi)_{n}&=\sum_{n,m}A_{n,m}\hat{L}_{n,m}, \\
\hat{L}_{n,m}&=-\frac{i}{2}\left(\hat{x}^{n+1}\hat{p}^{m+1}+\hat{p}^{m+1}\hat{x}^{n+1}\right),
\end{align}
where $A_{n,m}$ is some appropriate factor.
All $\hat{L}(\phi)$ can be expanded as the linear combination of the
$\hat{L}_{n,m}$, namely, the algebra generated by $\hat{L}_{n,m}$ is a generalization of the Virasoro algebra 
and is called $w_{\infty}$ algebra \cite{key-5}.
\subsection{Time-dependent anharmonic oscillators}
The time-dependent anharmonic oscillators are given by,
\begin{align*}
\hat{H}&=\hat{H}_{0}+\frac{1}{2}\lambda(t)\hat{x}^{n+2}, \\
\hat{H}_{0}&=\frac{\hat{p}^{2}}{2}+\frac{1}{2}\omega_{0}\hat{x}^{2}.
\end{align*}
In the interaction picture, the Hamiltonian reads,

\begin{equation*}
\hat{H}_{I}=\frac{1}{2}\omega(t)\hat{x}_{I}^{n+2}.
\end{equation*}
In the same way as in the squeezing argument, the time evolution of
the state can be written as follows,
\begin{equation*}
|0(t)\rangle=\hat{U}(t,0)|0\rangle,\ \hat{U}(t,0)=Te^{i\int_{0}^{t}dt\frac{1}{2}\lambda(t)\hat{x}_{I}^{n+2}}.
\end{equation*}

As is similar to the squeezing case, the time evolution of the state is
given by

\begin{equation}
|0(t)\rangle=\hat{U}(t,0)|0\rangle=e^{\theta_{n}\hat{L}_{n}(n)}\cdots e^{\theta_{0}\hat{L}_{n}(0)}|0\rangle,
\end{equation}
where $\theta_{m}=2(m+\frac{1}{2})\Delta t,\ \hat{L}_{n}(m)=\hat{L}_{n}(2(m+\frac{1}{2})\Delta t)$.

Then, in time-dependent oscillators, the time evolution of the state
can be obtained by the successive application of the N-th squeezing
operators.

\section{Particle production}

Here we calculate the number of particle generated by N-th order squeezing. The
expected number of particles in the N-th order squeezed state $|\theta\rangle$
is written as
\begin{equation}
_{n}\langle\theta|\hat{N}|\theta\rangle_{n}=\langle0|e^{-\theta \hat{L}_{n}}\hat{N}e^{\theta \hat{L}_{n}}|0\rangle=\langle0|\hat{a}_{\theta}^{\dagger}\hat{a}_{\theta}|0\rangle,
\end{equation}
where $\hat{a}_{\theta}$ and $\hat{a}^\dagger_{\theta}$ are 
\begin{align}
    \left\{ \ 
    \begin{aligned}
&\hat{a}_{\theta}  &=&\ e^{-\theta \hat{L}_{n}}\hat{a}e^{\theta \hat{L}_{n}},\\
  &&=&\ \frac{1}{2}\hat{K}(n,\hat{x})\left(
     (\cosh{\hat{\Omega}(n,\hat{x})})\hat{a}+(\sinh\hat{\Omega}(n,\hat{x}))\hat{a}^{\dagger}\right) \\
 &&& +\frac{1}{2}\left(a\cosh{\hat{\Omega}(n,\hat{x})}   +\hat{a}^{\dagger}\sinh{\hat{\Omega}(n,\hat{x})}\right)\hat{K}(n,\hat{x}),\\
&\hat{a}_{\theta}^{\dagger} & =&\ e^{-\theta L_{n}}\hat{a}^{\dagger}e^{\theta L_{n}},\\
 &&=&\frac{1}{2}\hat{K}(n,\hat{x})\left((\cosh{\hat{\Omega}(n,\hat{x})})\hat{a}^{\dagger}+(\sinh{\hat{\Omega}(n,\hat{x})})\hat{a}\right)\\
 &&& +\frac{1}{2}\left(\hat{a}^{\dagger}\cosh{\hat{\Omega}(n,\hat{x})}+\hat{a}\sinh\hat{\Omega}(n,\hat{x})\right)\hat{K}(n,\hat{x}).
    \end{aligned}
    \right.
\end{align}
Here,
\begin{equation}
\hat{\Omega}(n,\hat{x})=\log(1+n\theta\hat{x}^n)^{-\frac{1}{n}-\frac{1}{2}},\  \hat{K}(n,\hat{x})=(1+n\theta x^n)^{\frac{1}{2}}.
\end{equation}
These are a generalization of the Bogolyubov transformation.

Using these equations, we obtain
\begin{equation}
\langle0|\hat{a}_{\theta}^{\dagger}\hat{a}_{\theta}|0\rangle=\langle0|\left(\sinh\log(1+n(\theta \hat{x}^{n}))^{-\frac{1}{n}-\frac{1}{2}}\right)^{2}\hat{a}\hat{a}^{\dagger}|0\rangle.
\end{equation}
As a result, $_{n}\langle\theta|\hat{N}|\theta\rangle_{n}$ is given by
\begin{equation}
_{n}\langle\theta|\hat{N}|\theta\rangle_{n}=A_{0}\int dxe^{-x^{2}}\left(\sinh\log(1+n(\theta x^{n}))^{-\frac{1}{n}-\frac{1}{2}}\right)^{2},
\end{equation}
where $A_{0}=\left(\frac{\omega_{0}}{\pi}\right)^{1/4}.$
If $n=0$, $_{n}\langle\theta|\hat{N}|\theta\rangle_{n}$ reproduces
the result of the usual second-order squeezing,
\begin{align*}
_{0}\langle\theta|\hat{N}|\theta\rangle_{0}=\left(\sinh\theta\right)^{2}.
\end{align*}
In the case of $n\neq0$, if $\theta$ is small, we can expand $_{n}\langle\theta|\hat{N}|\theta\rangle_{n}$
in $\theta$ and the following equation is obtained,
\begin{align}
_{n}\langle\theta|\hat{N}|\theta\rangle_{n}&\sim\frac{1}{4}\theta^{2}(n+2)^{2}\Gamma\left(n+\frac{1}{2}\right) \notag \\
&-\frac{1}{8}\theta^{3}\left((-1)^{n}+1\right)n(n+2)^{2}\Gamma\left(\frac{3n}{2}+\frac{1}{2}\right)+O(\theta^{4}). \label{eq:NsqNum}
\end{align}
If the potential is $x^{4}$, namely for $n=2,$ $_{2}\langle\theta|\hat{N}|\theta\rangle_{2}$
reads
\begin{equation}
_{2}\langle\theta|\hat{N}|\theta\rangle_{2}=\frac{\sqrt{2}\pi e^{\frac{1}{2\theta}}(\theta-1)\text{erfc}\left(\frac{1}{\sqrt{2}\sqrt{\theta}}\right)+2\sqrt{\pi}\sqrt{\theta}(2\theta(\theta+1)(3\theta-1)+1)}{16\theta^{3/2}},
\end{equation}
where $\text{erfc}(x)$ is called the complementary error function,
given by

\begin{equation*}
\text{erfc}(x)=\frac{2}{\sqrt{\pi}}\int_{x}^{\infty}e^{-t^{2}}dt=\frac{e^{-x^{2}}}{x\sqrt{\pi}}\sum_{n=0}^{\infty}(-1)^{n}\frac{(2n)!}{n!(2x)^{2n}}.
\end{equation*}

If the potential is $x^{6}$, $n=4,$ $_{4}\langle\theta|\hat{N}|\theta\rangle_{4}$
becomes

\begin{align}
_{4}\langle\theta|\hat{N}|\theta\rangle_{4} & =\frac{\sqrt{\pi}}{64\theta} \notag \\
&\left(-10\pi\theta^{3/2}J_{-\frac{1}{4}}\left(\frac{1}{4\sqrt{\theta}}\right){}^{2}-2\pi\sqrt{\theta}\left(60\theta^{2}+9\theta-2\right)\left(J_{\frac{1}{4}}\left(\frac{1}{4\sqrt{\theta}}\right)\right)^{2} \right. \notag \\
&-2\sqrt{2}\pi\sqrt{\theta}J_{-\frac{1}{4}}\left(\frac{1}{4\sqrt{\theta}}\right)\left(\left(-30\theta^{2}-3\theta+1\right)J_{\frac{1}{4}}\left(\frac{1}{4\sqrt{\theta}}\right)\right) \notag \\
&-2\sqrt{2}\pi\sqrt{\theta}J_{-\frac{1}{4}}\left(\frac{1}{4\sqrt{\theta}}\right)\left((1-15\theta)\sqrt{\theta}Y_{-\frac{3}{4}}\left(\frac{1}{4\sqrt{\theta}}\right)\right) \notag \\
&-32\theta+2\pi\sqrt{\theta}(15\theta+1)\left(J_{\frac{3}{4}}\left(\frac{1}{4\sqrt{\theta}}\right)\right)^{2}+2\pi\sqrt{\theta}(15\theta+1)\left(J_{\frac{5}{4}}\left(\frac{1}{4\sqrt{\theta}}\right)\right)^{2} \notag \\
&+12\pi\theta(5\theta+1)J_{-\frac{3}{4}}\left(\frac{1}{4\sqrt{\theta}}\right)J_{\frac{1}{4}}\left(\frac{1}{4\sqrt{\theta}}\right) \notag \\
&+\sqrt{2}\pi J_{-\frac{3}{4}}\left(\frac{1}{4\sqrt{\theta}}\right)\left(2\sqrt{\theta}(15\theta+1)J_{\frac{3}{4}}\left(\frac{1}{4\sqrt{\theta}}\right)-Y_{-\frac{1}{4}}\left(\frac{1}{4\sqrt{\theta}}\right)\right) \notag \\
&\left. -\sqrt{2}\pi J_{-\frac{5}{4}}\left(\frac{1}{4\sqrt{\theta}}\right)\left(2(1-15\theta)\theta J_{\frac{1}{4}}\left(\frac{1}{4\sqrt{\theta}}\right)-Y_{\frac{1}{4}}\left(\frac{1}{4\sqrt{\theta}}\right)\right)\right),
\end{align}
where $J_{a}$ and $Y_{a}$ are
Bessel functions of the first and second kind.

\begin{figure}[]
    \begin{minipage}{0.5\hsize}
        \begin{center}
            \includegraphics[scale=0.3]{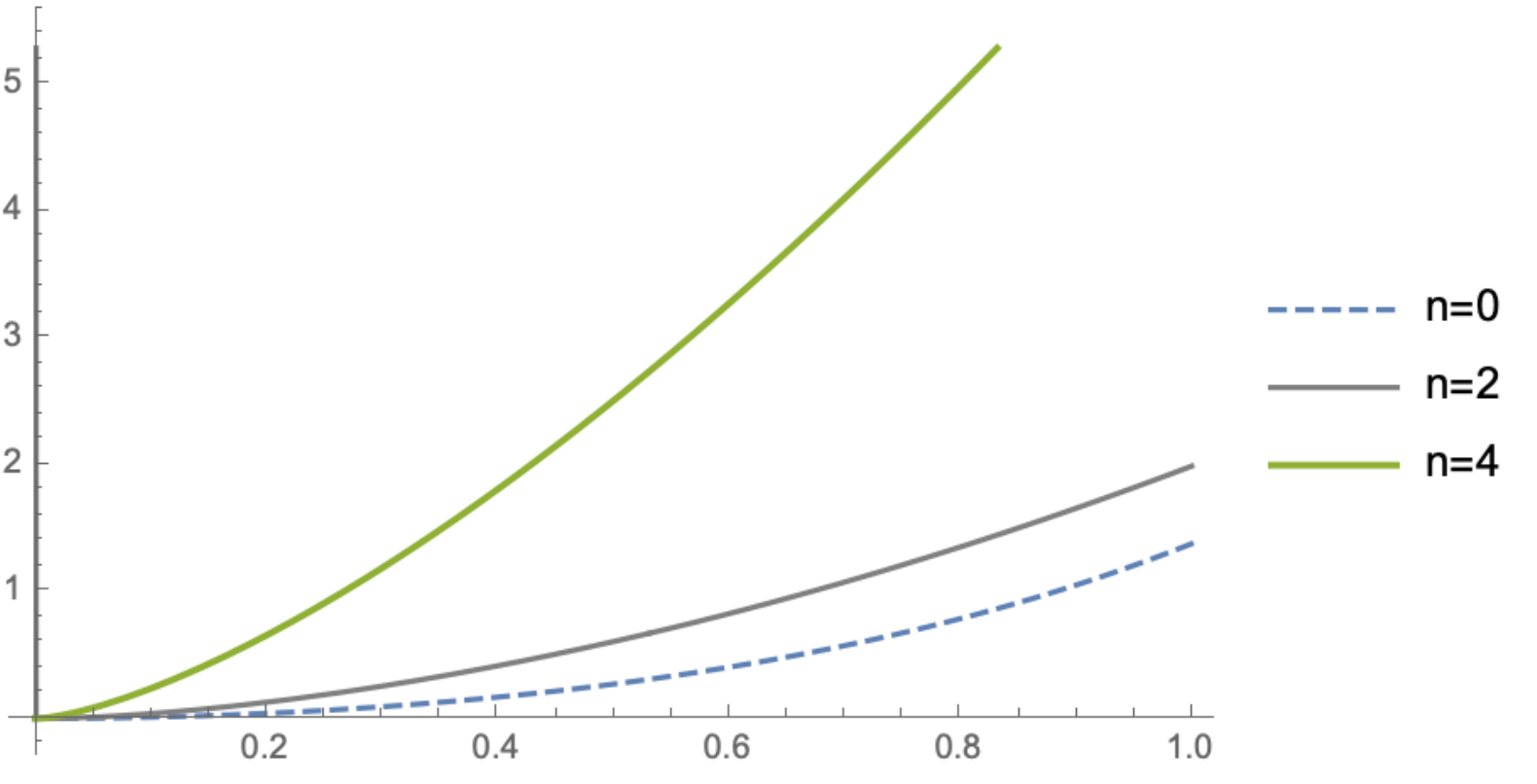}
        \end{center}
        \subcaption{$0\leq\theta\leq1$}
    \end{minipage}
    \begin{minipage}{0.5\hsize}
        \begin{center}
            \includegraphics[scale=0.3]{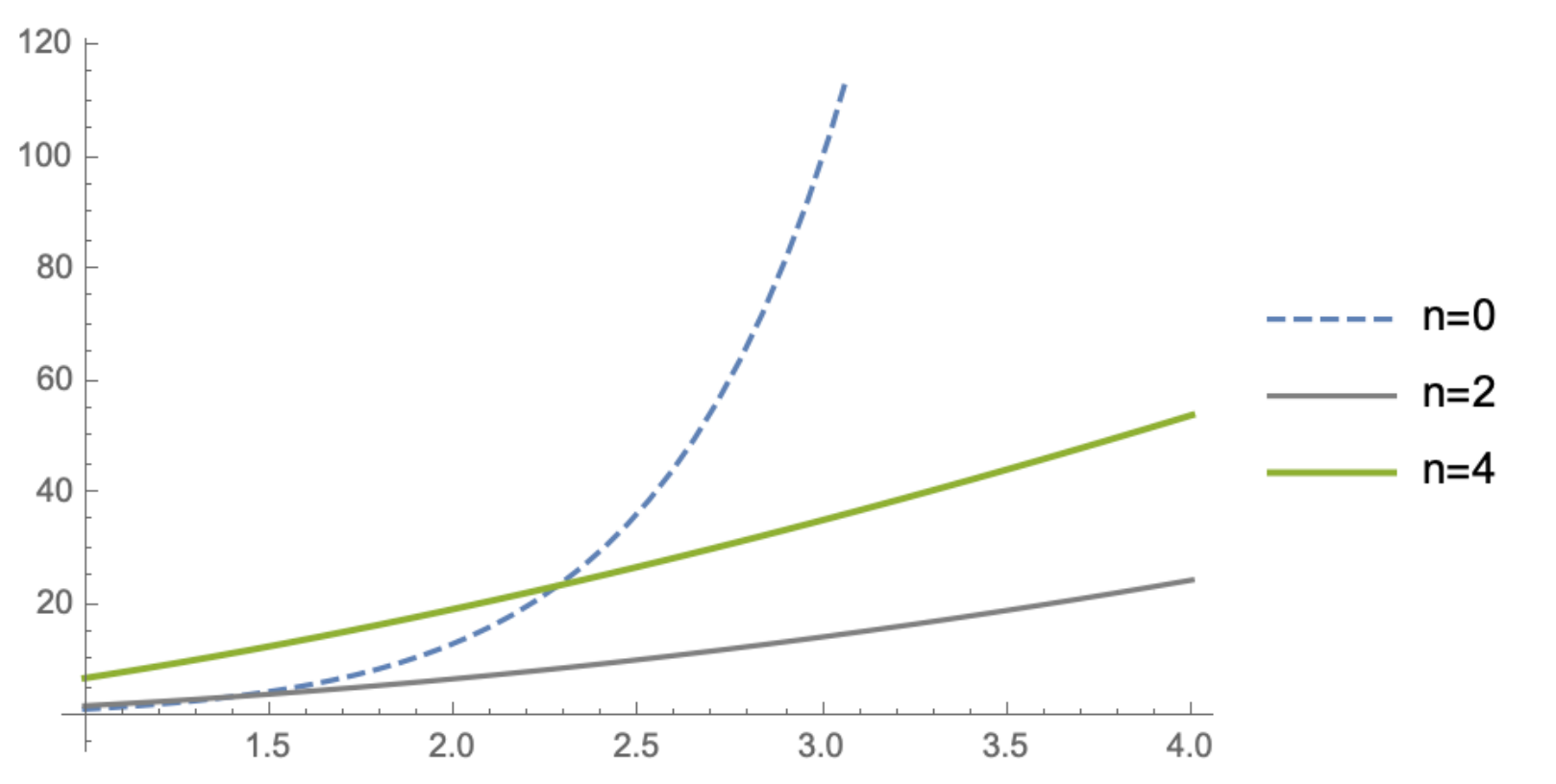}
        \end{center}
        \subcaption{$1\leq\theta\leq4$}
    \end{minipage}
    \caption{\label{fig:particle-number-generation}Particle number generation
    of N-th order squeezing. The horizontal axis represents squeezing parameter $\theta$
    and the vertical axis represents particle number.
    The line of $n=0$ cross thst of $n=2$ at $\theta = 1.39251$ and cross that of $n=4$ at $\theta = 2.28158$. }    
\end{figure}

The number of particles produced in quantum mechanics with having the
time dependent $x^{n+2}$ potential for $n=0,2,4$ is plotted in fig.\ref{fig:particle-number-generation}.

This shows that for small $\theta$, the number of generated particles increase as
the order of squeezing becomes higher.

\section{The uncertainty relation of higher order}

The usual second-order squeezed state satisfied the minimum uncertainty
relation. We calculate the uncertainty relation in the
case of the N-th order squeezing.

First, to determine $_{n}\langle\theta|\Delta x^{2}|\theta\rangle_{n}=\ _{n}\langle\theta|x^{2}|\theta\rangle_{n}-{}_{n}\langle\theta|x|\theta\rangle_{n}^{2}$,
we note
\begin{equation}
\langle x_{-\theta}|\equiv\langle x|e^{\theta \hat{L}_{n}}
\end{equation}
is an eigenstate of $\hat{x}$ because of
\begin{equation*}
\langle x_{-\theta}|\hat{x}\equiv\langle x|e^{\theta \hat{L}_{n}}\hat{x}=\langle x|(1+n(-\theta\hat{x}^{n}))^{-\frac{1}{n}}\hat{x}e^{\theta \hat{L}_{n}}=\langle x_{-\theta}|(1+n(-\theta \hat{x}^{n}))^{-\frac{1}{n}}\hat{x}.
\end{equation*}
Then, we obtain
\begin{equation}
\langle x|\theta\rangle_{n}=\langle x_{-\theta}|0\rangle=A_{0}e^{-\frac{1}{2}\omega_{0}(1+n(\theta x^{n}))^{-\frac{2}{n}}x^{2}}.
\end{equation}
Using this vacuum state, we can calculate the uncertainty relation
perturbatively in $\theta$.

As a result, if $n$ is even, we obtain
\begin{align}
_{n}\langle\theta|\Delta \hat{x}^{2}|\theta\rangle_{n}&=|A_{0}|^{2}\left(\frac{\sqrt{\pi}}{2\omega_{0}^{3/2}}+2\Gamma(\frac{5+n}{2})\omega_{0}^{-\frac{3}{2}-\frac{n}{2}}\theta+(3+n)\Gamma(\frac{5}{2}+n)\omega_{0}^{-2-n}\theta^{2}\right) \notag \\
&+O(\theta^{3}),
\end{align}
while if $n$ is odd, we have
\begin{align}
_{n}\langle\theta|\Delta \hat{x}^{2}|\theta\rangle_{n}&=|A_{0}|^{2}\left(\frac{\sqrt{\pi}}{2\omega_{0}^{3/2}}+(3+n)\Gamma(\frac{5}{2}+n)\omega_{0}^{-2-n}\theta^{2}\right) \notag \\
&-4|A_{0}|^{4}\Gamma(2+\frac{n}{2})^{2}\omega_{0}^{-2-n}\theta^{2}+O(\theta^{3}).
\end{align}
In estimating $_{n}\langle\theta|\Delta \hat{p}^{2}|\theta\rangle_{n}$
, we will make use of $\langle x|\theta\rangle_{n}$ \\ 
and $_{n}\langle\theta|p|\theta\rangle_{n}=\int dx{}_{n}\langle\theta|x\rangle i\frac{\partial}{\partial x}\langle x|\theta\rangle_{n}$.

As a result, if $n$ is even, we get
\begin{align}
_{n}\langle\theta|\Delta \hat{p}^{2}|\theta\rangle_{n}&=|A_{0}|^{2}\left(\frac{\sqrt{\pi}\sqrt{\omega_{0}}}{2}+(1+n)\Gamma(\frac{3+n}{2})\omega_{0}^{\frac{1}{2}-\frac{n}{2}}\theta\right), \notag \\
&+|A_{0}|^{2}\left(\frac{1}{2}(1+n)\Gamma(\frac{3}{2}+n)\omega_{0}^{\frac{1}{2}-n}\theta^{2}\right)+O(\theta^{3}),
\end{align}
and if $n$ is odd, we have
\begin{align}
_{n}\langle\theta|\Delta \hat{p}^{2}|\theta\rangle_{n}=|A_{0}|^{2}\left(\frac{\sqrt{\pi}\sqrt{\omega_{0}}}{2}+\frac{1}{2}(1+n)\Gamma(\frac{3}{2}+n)\omega_{0}^{\frac{1}{2}-n}\theta^{2}\right)+O(\theta^{3}).
\end{align}

From $_{n}\langle\theta|\Delta \hat{x}^{2}|\theta\rangle_{n}$ and $_{n}\langle\theta|\Delta \hat{p}^{2}|\theta\rangle_{n}$,
the uncertainty relation of N-th order squeezing
is given as follows.
If $n$ is even,
\begin{align}
_{n}\langle\theta|\Delta \hat{x}^{2}|\theta\rangle_{n}{}_{n}\langle\theta|\Delta p^{2}|\theta\rangle_{n} & =\frac{1}{4}-\frac{1}{\sqrt{\pi}}\Gamma\left(\frac{n+3}{2}\right)\text{\ensuremath{\omega_{0}}}^{-\frac{n}{2}}\theta \notag \\
 & +\frac{1}{2\sqrt{\pi}}\left(n^{2}+5n+5\right)\Gamma\left(n+\frac{3}{2}\right)\text{\ensuremath{\omega_{0}}}^{-n}\theta^{2}  \\
 & -\frac{2}{\pi}(n+1)\Gamma\left(\frac{n+3}{2}\right)\Gamma\left(\frac{n+5}{2}\right)\text{\ensuremath{\omega_{0}}}^{-n}\theta^{2}+O\left(\theta^{3}\right). \notag
\end{align}
If $n$ is odd,
\begin{align}
_{n}\langle\theta|\Delta x^{2}|\theta\rangle_{n}{}_{n}\langle\theta|\Delta p^{2}|\theta\rangle_{n} & =\frac{1}{4} \notag \\
 & -\frac{2}{\pi}\Gamma\left(\frac{n+4}{2}\right)^{2}\text{\ensuremath{\omega_{0}}}^{-n}\theta^{2}  \\
 & +\frac{1}{2\sqrt{\pi}}\left(n^{2}+5n+5\right)\Gamma\left(n+\frac{3}{2}\right)\text{\ensuremath{\omega_{0}}}^{-n}\theta^{2}+O\left(\theta^{3}\right). \notag
\end{align}

As an example, if we take $n=2,$ the uncertainty
relation is

\begin{align}
_{2}\langle\theta|\Delta x^{2}|\theta\rangle_{2}{}_{2}\langle\theta|\Delta p^{2}|\theta\rangle_{2} & =\frac{1}{4}-\frac{3}{4}\frac{\theta}{\omega_{0}}+\frac{75}{8}\frac{\theta^{2}}{\omega_{0}^{2}}+O(\theta^{3}).
\end{align}

The above results indicate that the minimum uncertainty relation
is broken at the low order of perturbation in $\theta$ in the N-th order
squeezing.

\section{Phase space of the higher order}

We introduce a Husimi function \cite{key-16} to
determine the structure of phase space. The Husimi function is defined
by coherent representation,
\begin{equation*}
H(\alpha)=\langle\alpha|\hat{\rho}|\alpha\rangle,
\end{equation*}
where 
\begin{equation*}
|\alpha\rangle=A_{\alpha}e^{\alpha \hat{a}^{\dagger}}|0\rangle,A_{\alpha}=e^{-\frac{|\alpha|^{2}}{2}}.
\end{equation*}
For the vacuum state, $\hat{\rho}=|0\rangle\langle0|$,
the Husimi function $H_{0}(\alpha)$ is

\begin{equation*}
H_{0}(\alpha)=e^{-|\alpha|^{2}}.
\end{equation*}

To get the N-th order squeezed Husimi function $H_{\theta_{n}}(\alpha)=\langle\bar{\alpha}|\theta_{n}\rangle\langle\theta_{n}|\alpha\rangle$,
we note
\begin{equation}
_{n}\langle\theta|\alpha\rangle=\int\frac{d\beta d\bar{\beta}}{\pi}e^{-\theta\mathcal{L}_{n}}\langle0|\beta\rangle\langle\bar{\beta|}\alpha\rangle=e^{-\theta\mathcal{L}_{n}}\langle0|\alpha\rangle,
\end{equation}

\begin{align*}
\mathcal{L}_{n}&=\frac{1}{2}\left(\sqrt{\frac{\omega_{0}}{2}}\right)\left(\sqrt{\frac{1}{2\omega_{0}}}\right)^{n+1} \\
&\left((\alpha+\frac{\partial}{\partial\alpha}+\frac{\text{\ensuremath{\bar{\alpha}}}}{2})^{n+1}(\alpha-\frac{\partial}{\partial\alpha}-\frac{\text{\ensuremath{\bar{\alpha}}}}{2})+(\alpha-\frac{\partial}{\partial\alpha}-\frac{\text{\ensuremath{\bar{\alpha}}}}{2})(\alpha+\frac{\partial}{\partial\alpha}+\frac{\text{\ensuremath{\bar{\alpha}}}}{2})^{n+1}\right)
\end{align*}
where we have used the following formula,
\begin{equation*}
\langle\bar{\beta}|f(a,a^{\dagger})|\alpha\rangle=f(\alpha,\frac{\partial}{\partial\alpha}+\frac{\bar{\alpha}}{2})\langle\bar{\beta}|\alpha\rangle.
\end{equation*}
Because
\begin{equation*}
\langle0|\alpha\rangle=e^{-\frac{|\alpha|^{2}}{2}},
\end{equation*}
the term $\frac{\partial}{\partial\alpha}+\frac{\text{\ensuremath{\bar{\alpha}}}}{2}$
vanishes and $_{n}\langle\theta|\alpha\rangle$ becomes
\begin{equation}
_{n}\langle\theta|\alpha\rangle=e^{-\theta\mathcal{L}'_{n}}e^{-\frac{|\alpha|^{2}}{2}},
\end{equation}
\begin{equation*}
\mathcal{L}'_{n}=\sqrt{\frac{\omega_{0}}{2}}\left(\sqrt{\frac{1}{2\omega_{0}}}\right)^{n+1}\alpha{}^{n+2}.
\end{equation*}

Therefore, the N-th order squeezed Husimi function
is given by
\begin{equation}
H_{\theta_{n}}(\alpha)=e^{-\theta\sqrt{\frac{\omega_{0}}{2}}\left(\sqrt{\frac{1}{2\omega_{0}}}\right)^{n+1}\left(\alpha^{n+2}+\bar{\alpha}^{n+2}\right)}e^{-|\alpha|^{2}}.\label{eq:husimi_n}
\end{equation}

\begin{figure}[p]
    \begin{minipage}{0.5\hsize}
        \begin{center}
            \includegraphics[scale=0.3]{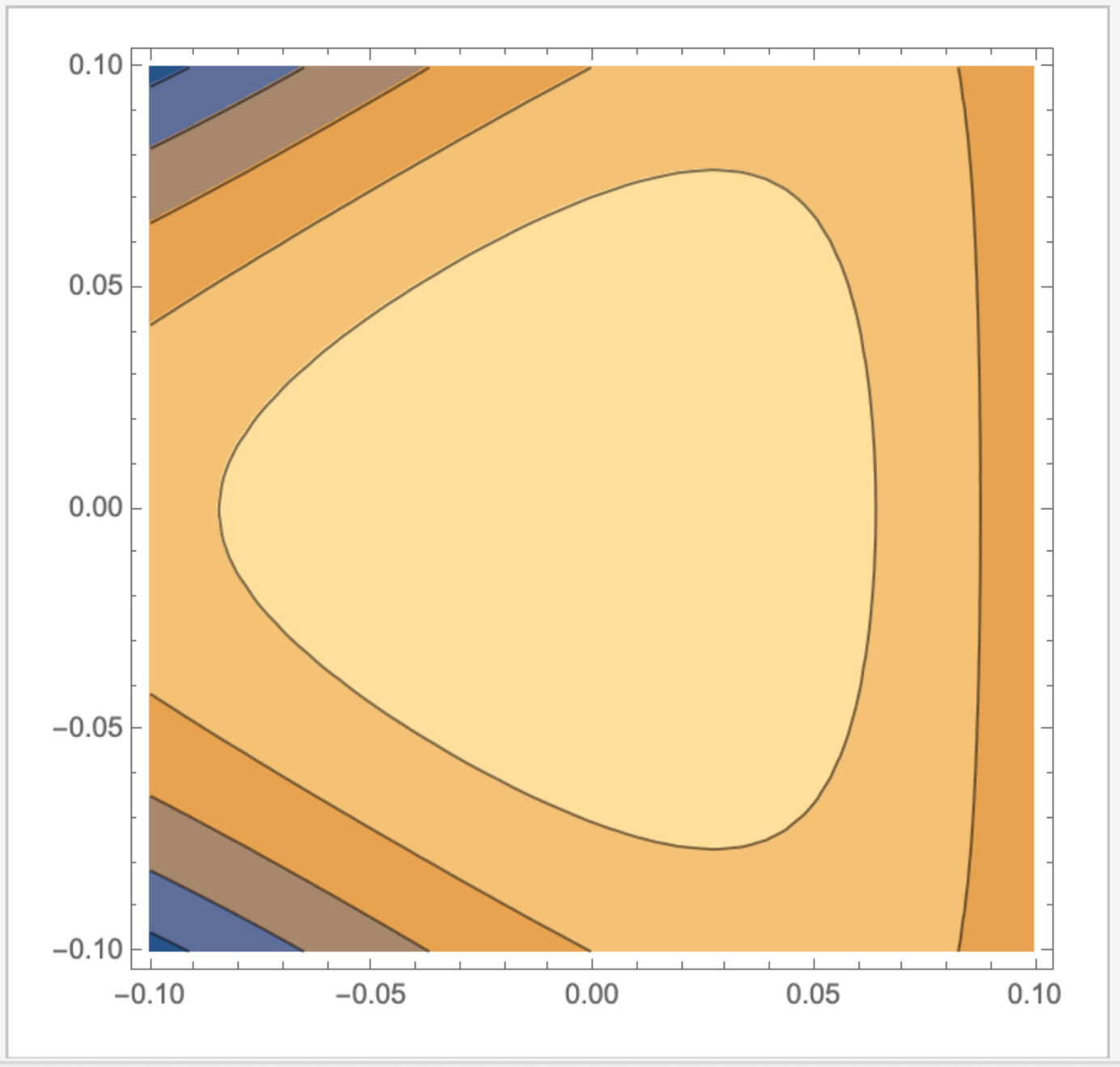}
        \end{center}
        \subcaption{$n=1$}
    \end{minipage}
    \begin{minipage}{0.5\hsize}
        \begin{center}
            \includegraphics[scale=0.3]{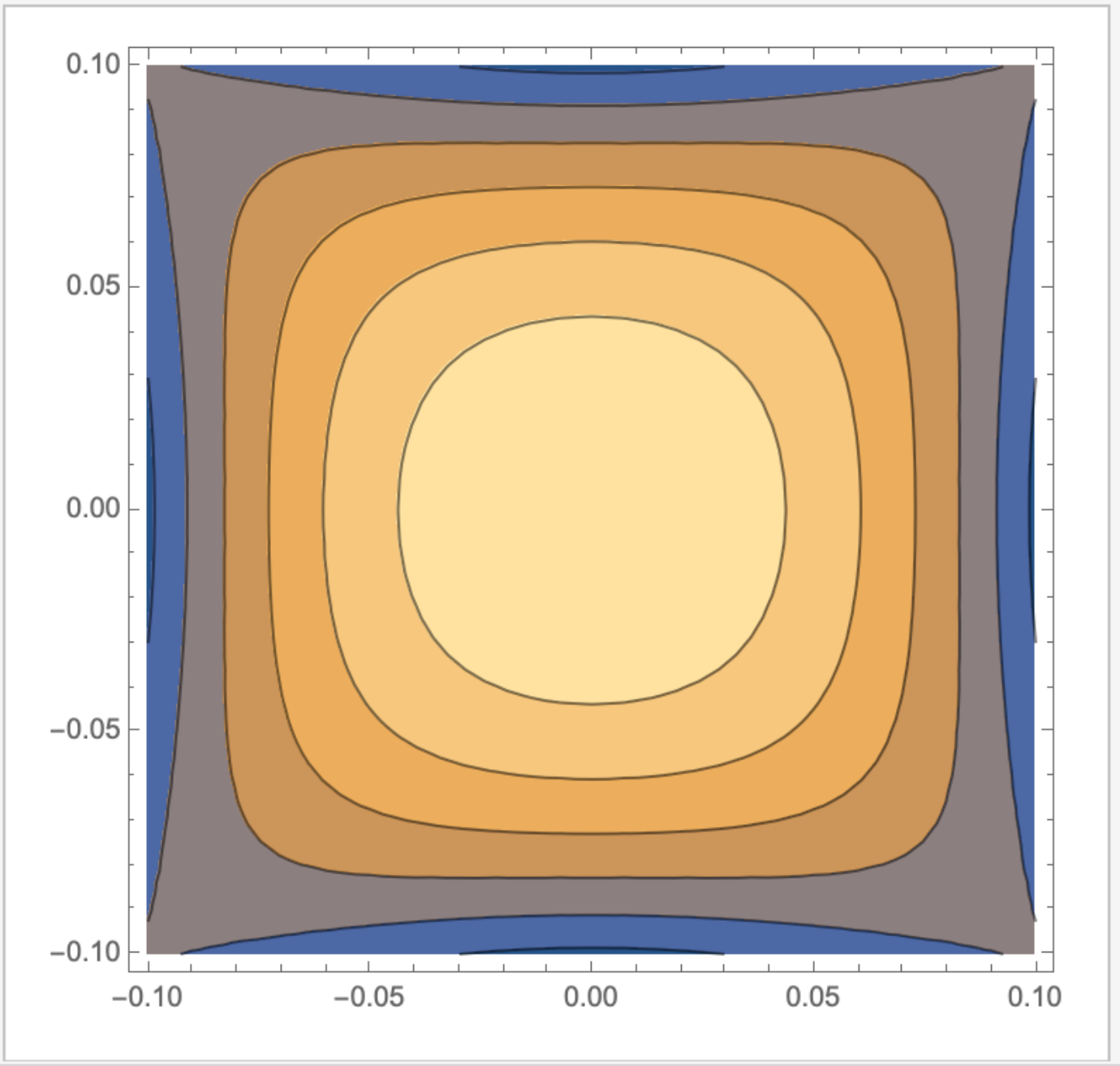}
        \end{center}
        \subcaption{$n=2$}
    \end{minipage}
    \begin{minipage}{0.5\hsize}
        \begin{center}
            \includegraphics[scale=0.3]{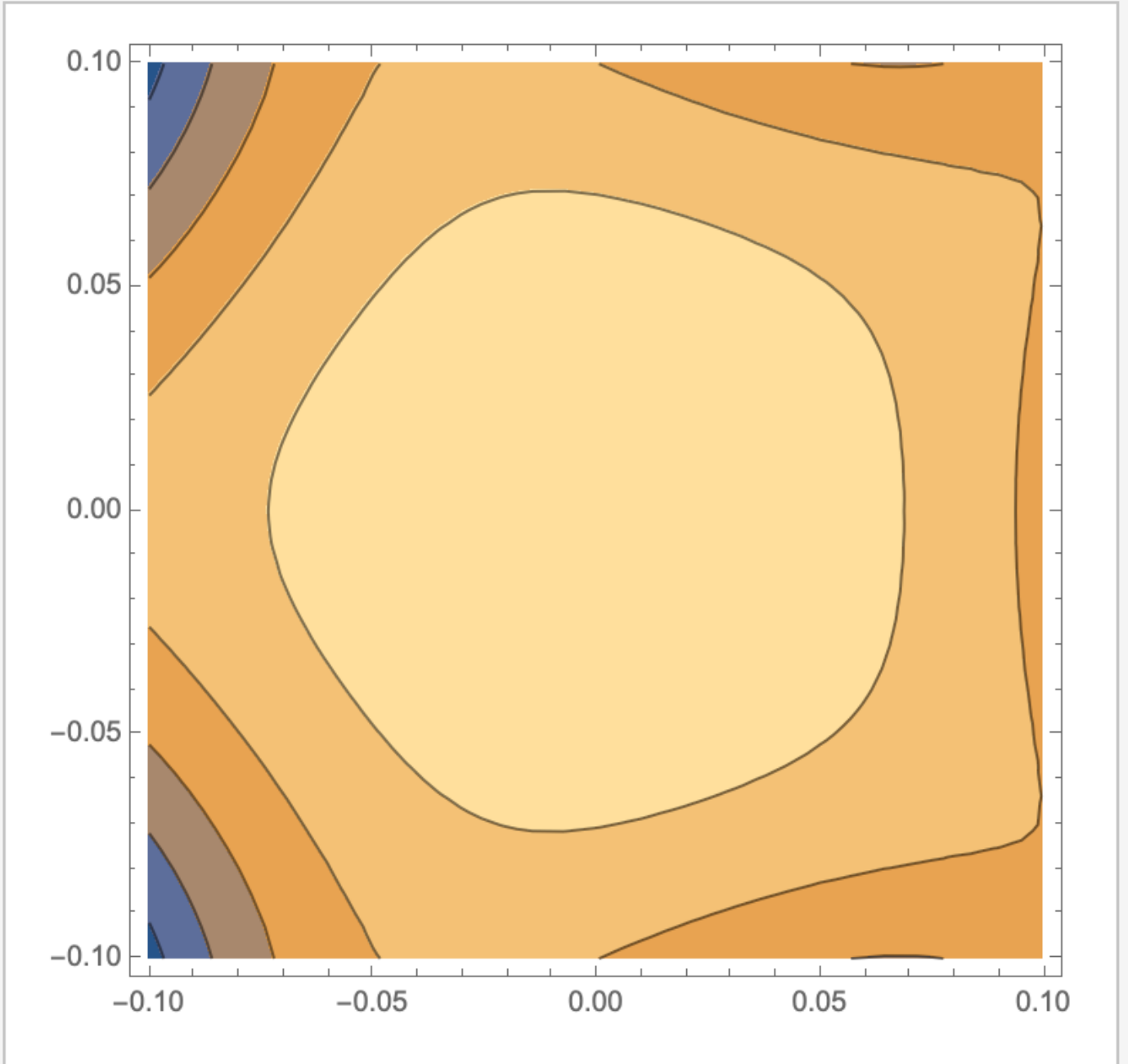}
        \end{center}
        \subcaption{$n=3$}
    \end{minipage}
    \begin{minipage}{0.5\hsize}
        \begin{center}
            \includegraphics[scale=0.3]{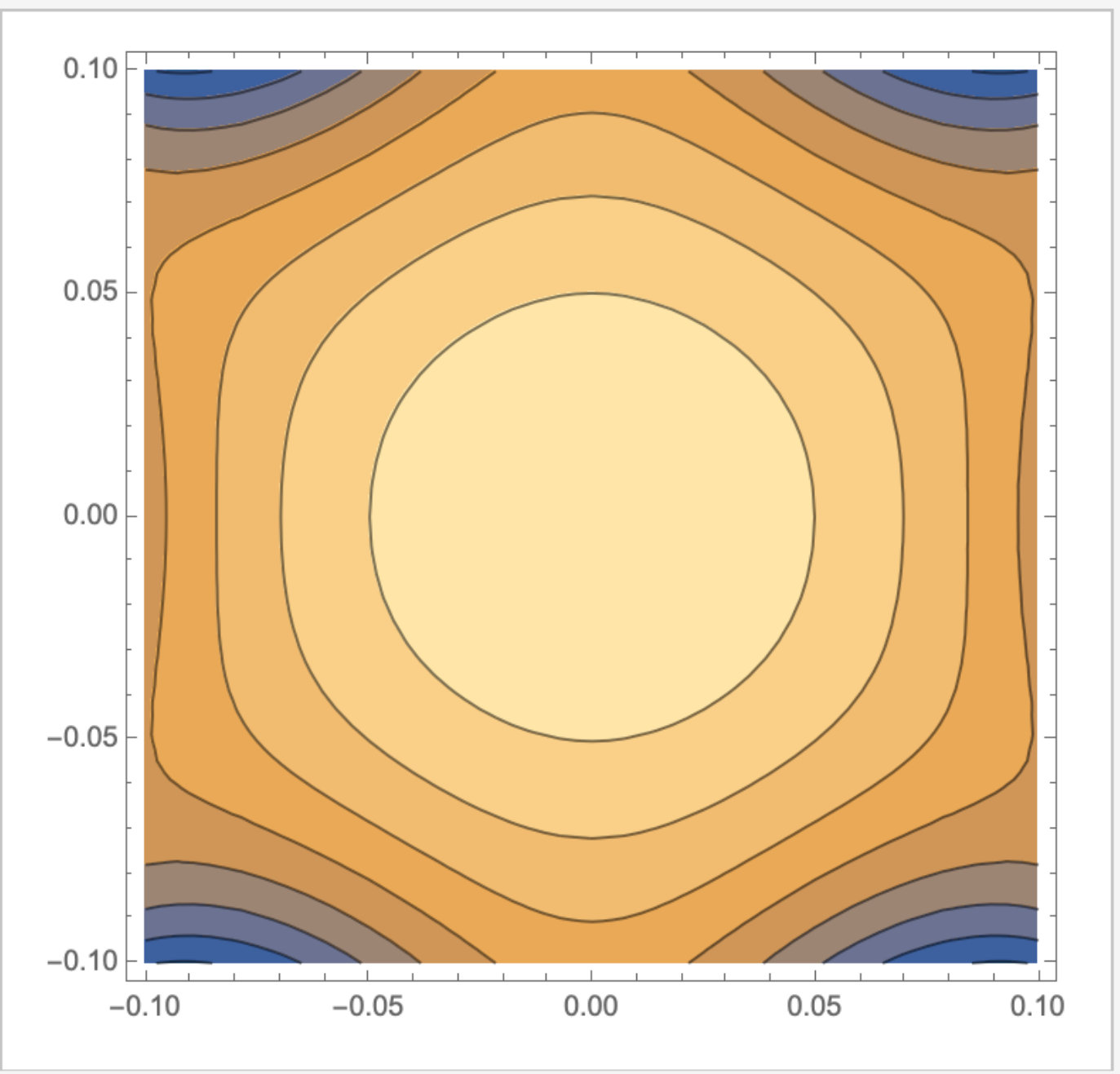}
        \end{center}
        \subcaption{$n=4$}
    \end{minipage}
\caption{\label{fig:Phase-Space}
Behavior of Husimi function (\ref{eq:husimi_n}) in phase space for $n=1, \dots, 4$, where $n$ characterizes the power of the quantum-mechanical potential $x^{2+n}$. The squeezing parameter $\theta$ is set to $\theta=0.5$, and the characteristic frequency $\omega_0$ of the free theory is set to $\omega_0 = 0.01$. The horizontal axis represents position $x$ and the vertical axis represents momentum $p$.}
\end{figure}

As a final result, the contour lines of phase space are depicted
in fig. \ref{fig:Phase-Space}. From the figure, we have found that
the N-th order squeezing is narrowed in proportion to $(2+n)$-th
square root.

\section{Discussion}

As the beyond squeezing effect, we have proposed the N-th order squeezing
based on the Virasoro algebra, in which the N-th order squeezing is
induced by the N-th order time-dependent potential. That is, the usual
second-order squeezing operator is generalized to the higher-order Virasoro operators.
The $\hat{x}$ and $\hat{p}$ are subject to local scale
deformation in the opposite direction also in N-th order squeezing.

We have obtained a formula of the particle number produced by the
N-th order squeezing, when the squeezing parameter $\theta$ is small. The
formula implies that the higher the order of the squeezing is, the
larger the number of generated particle is.

It is interesting to note that the extension of N-th order squeezing
in any direction is related to the $W_{\infty}$-algebra. The algebra
of the transformation that preserves the area of phase space\footnote{This is called $\mathrm{sdiff}(\Sigma)$ where $\Sigma$ is phase space of $(x,p)$ .}
is called $w_{\infty}$-algebra, and the $W_{\infty}$-algebra is its
the quantized version in which the area conservation is not always preserved. These
algebras represent deep connections with integrable systems \cite{key-13},
and the relationship between Nth-order squeezing states and integrable
systems need to be further studied.

In the context of string theory, a similar Virasoro algebra and $W_{\infty}$-algebra
with $x$ and $p$ in the $AdS_{2}/CFT_{1}$ correspondence appears
on the $CFT_{1}$ side \cite{key-14}. 
Particles near the horizon of an extremal Reissner-Nordstrom black hole is described by conformal
mechanics \cite{key-15}. In $AdS_{2}/CFT_{1}$, the algebra of
symmetry is the Virasoro algebra, and it is possible that understanding
of the relationship between the Virasoro algebra in $AdS_{2}/CFT_{1}$
and that in the N-th order squeezing can elucidate the role of N-th order
squeezing in black holes.

In the theory of reheating after inflation,
 the effect of squeezing by parametric resonance has been attracting attention 
 in connection with the problem of baryogenesis \cite{key-17},
  and it may be applied as an effective model for the N-th order squeezing.

The N-th order squeezing obtained in this discussion can be generalized to the two-mode Bogolyubov transformation or fermionic version. 
It is interesting to investigate the physics, which might bring the extensions of well-known theories such as Bardeen-Cooper-Schrieffer theory \cite{key-18}.

These studies will be the subject in the near future.

\section*{Acknowledgments}

We are deeply grateful to Shiro Komata for discussions which clarify
important points on Virasoro algebra. We are indebted to Ken Yokoyama
for reading this paper and giving useful comments. We would also like
to thank Naoaki Fujimoto and Noriaki Aibara for reading this paper.

\end{document}